\documentclass[amsmath,amssymb,superscriptaddress]{revtex4}
\usepackage{amsmath}
\usepackage{graphicx}
\usepackage{rotating}
\begin{document}
\hspace*{5 in}CUQM - 139
\title{Study of the generalized quantum isotonic nonlinear oscillator potential}
\author{Nasser Saad}
\email{nsaad@upei.ca}
\affiliation{Department of Mathematics and Statistics,
University of Prince Edward Island,
550 University Avenue, Charlottetown,
PEI, Canada C1A 4P3}
\author{Richard L. Hall}
\email{rhall@mathstat.concordia.ca}
\affiliation{Department of Mathematics and Statistics, Concordia University,
1455 de Maisonneuve Boulevard West, Montr\'eal,
Qu\'ebec, Canada H3G 1M8}
\author{Hakan \c{C}ift\c{c}i}
\email{hciftci@gazi.edu.tr}
\affiliation{Department of Physics, Faculty of Arts and Sciences,
Gazi University, 06500 Ankara, Turkey.}
\author{\"Ozlem Ye\c{s}ilta\c{s}}
\email{yesiltas@gazi.edu.tr}
\affiliation{Department of Physics, Faculty of Arts and Sciences,
Gazi University, 06500 Ankara, Turkey.}


\def\dbox#1{\hbox{\vrule  
                        \vbox{\hrule \vskip #1
                             \hbox{\hskip #1
                                 \vbox{\hsize=#1}%
                              \hskip #1}%
                         \vskip #1 \hrule}%
                      \vrule}}
\def\qed{\hfill \dbox{0.05true in}}  
\def\square{\dbox{0.02true in}} 
\begin{abstract}
\noindent We study the generalized quantum isotonic oscillator Hamiltonian given by  $H=-d^2/dr^2+l(l+1)/r^2+w^2r^2+2g(r^2-a^2)/(r^2+a^2)^2$, $g>0$. Two approaches are explored. A method for finding the quasi-polynomial solutions is presented, and explicit expressions for these polynomials are given, along with the conditions on the potential parameters.  By using the asymptotic iteration method we show how  the eigenvalues of this Hamiltonian for arbitrary values of the parameters $g, w$ and $a$ may be found to high accuracy. 
\end{abstract}
\maketitle
\noindent {\bf keyword:} Non-linear oscillators; Non-polynomial potentials; Gol'dman and Krivchenkov potential; Asymptotic Iteration Method; Quantum integrable systems; Laguerre polynomials.\\


\section{Introduction}

\noindent Recently, Cari\~nena \emph{et al.}~\cite{Carinena} studied a quantum nonlinear oscillator potential whose Schr\"odinger equation reads
\begin{equation}\label{eq1}
\left[-{d^2\over dx^2}+x^2+8{2x^2-1\over (2x^2+1)^2}\right]\psi_n(x)=E_n\psi(x)
\end{equation}
The interest in this problem came from the fact that it is exactly solvable, in a sense that the exact eigenenergies and eigenfunctions can be obtained explicitly. Indeed, Cari\~nena \emph{et al.} \cite{Carinena} were able to show that
\begin{equation}\label{eq2}
\left\{ \begin{array}{l}
 \psi_n(x)={P_n(x)\over (2x^2+1)}e^{-x^2/2},\\ \\
  E_n=-3+2n,\quad n=0,3,4,5,\dots
       \end{array} \right.
\end{equation}
where the polynomials factors $P_n(x)$ are related to the Hermite polynomials by means of 
\begin{equation}\label{eq3}
P_n(x)=\left\{ \begin{array}{ll}
 1 &\mbox{ if $n=0$} \\
  H_n(x)+4nH_{n-2}(x)+4n(n-3)H_{n-4}(x) &\mbox{ if $n=3,4,5,\dots$}
       \end{array} \right.
\end{equation} 
In a more recent work,  Fellows and Smith \cite{fellows} showed that the potential $V(x)=x^2+8(2x^2-1)/(2x^2+1)^2$  as well as, for certain values of the parameters $w$, $g$ and $a$, the potential $V(x)=w^2x^2+2g{(x^2-a^2)/(x^2+a^2)^2}$ of the Schr\"odinger equation
\begin{equation}\label{eq4}
\left[-{d^2\over dx^2}+w^2x^2+2g{x^2-a^2\over (x^2+a^2)^2}\right]\psi_n(x)=2E_n\psi(x),
\end{equation}
 are indeed supersymmetric partners of the harmonic oscillator potential. Using the supersymmetric approach,  the authors were able to construct an infinite set of exact soluble potentials, along with their eigenfunctions and eigenvalues. Very recently, Sesma \cite{sesma}, using a M\"obius transformation,  was able to transform Eq.(\ref{eq4}) into a confluent Heun equation \cite{heun} and thereby obtain an efficient algorithm to solve the Schr\"odinger equation (\ref{eq4}) numerically.  
\vskip0.1true in
\noindent The purpose of the present work is to provide a detailed solution, by means of the quasi-polynomial solutions and the application of the asymptotic iteration method \cite{brodie,C3,ciftci1,ciftci2}, for the Schr\"odinger equation
\begin{equation}\label{eq5}
\left[-{d^2\over dr^2}+{l(l+1)\over r^2}+w^2r^2+2g{r^2-a^2\over (r^2+a^2)^2}\right]\psi(r)=2E\psi(r), 
\end{equation}
where $l$ is the  angular momentum number $l=-1,0,1,\dots$. Our results show that the quasi-exact solutions of Sesma \cite{sesma} as well the results of Cari\~nena \emph{et al.} \cite{Carinena}  follow as special cases of our general approach.  The present article is organized as follows. In the next section, some preliminary analysis of the Schr\"odinger equation (\ref{eq5}) is presented. A general approach for finding polynomial solutions of Eq.(\ref{eq5}), for certain values of parameters $w$ and $g$, is presented, and is based on a recent work of Ciftci \emph{et al.} \cite{C3} for solving the second-order linear differential equation
\begin{equation}\label{eq6}
\left(\sum_{i=0}^3 a_{3,i}x^i\right) y''+\left(\sum_{i=0}^2 a_{2,i}x^i\right) y'-\left(\sum_{i=0}^1 \tau_{1,i}x^i\right) y=0.
\end{equation}
 More general quasi-exact solutions, including the results of Sesma \cite{sesma},  are discussed in section III. Unrestricted solutions of Eq.(\ref{eq5}) based on the asymptotic iteration method are discussed in Section IV.
\section{Generalized quantum isotonic oscillator - preliminary results} 
\noindent A simple scaling argument, using $r=a^2 x$, allows us to write the equation (\ref{eq5}) as
\begin{equation}\label{eq7}
\left[-{d^2\over dx^2}+{l(l+1)\over x^2}+(wa^2)^2x^2+2g{x^2-1\over (x^2+1)^2}\right]\psi(x)=2Ea^2\psi(x).
\end{equation}
A further substitution $z=x^2+1$ yields a differential equation with two regular singular points at $z=0,1$ and one irregular singular point of rank $2$ at $z=\infty$. The roots $\mu$'s of the indicial equation for the regular singular point $z=0$ reads $\mu_\pm={1\over 2}(1\pm\sqrt{1+4g})$, while the roots of the indicial equation at $z=1$ are $\mu_{+}=(l+1)/2$ and $\mu_{-}=-l/2$. Since the singularity for $z\rightarrow \infty$ corresponds to that for $x\rightarrow\infty$, it is necessary that the solution for $z\rightarrow \infty$ behave as
$\psi(x)\sim \exp(-wa^2x^2/2)$. Consequently, we may assume the general solution of equation (\ref{eq7}) which vanishes at the origin and at infinity takes the form 
\begin{equation}\label{eq8}
\psi_n(x)=x^{l+1}(x^2+1)^{\mu} e^{-{wa^2\over 2}x^2}f_n(x).
\end{equation}
A straightforward calculation shows that $f_n(x)$ are the solutions of the second-order homogeneous linear differential
equation
\begin{align}\label{eq9}
f''(x)&+\left({2(l+1)\over x}+{4\mu x\over x^2+1}-2wa^2x\right)f'(x)\notag\\
&+\left[2Ea^2-wa^2(2l+3+4\mu)+{2\mu(2l+3+2wa^2)+4\mu(\mu-1)-2g\over x^2+1}+{4(g-\mu(\mu-1))\over (x^2+1)^2}\right]f(x)=0.
\end{align}
In the next sections, we attempt to give a general solution of this equation. For now, we assume that $\mu$ takes the value of the indicial root
\begin{equation}\label{eq10}
\mu\equiv \mu_{-}={1\over 2}(1-\sqrt{1+4g})
\end{equation}
which allows us to write Eq.(\ref{eq9}) as
\begin{align}\label{eq11}
&f_n''(x)+\left({2(l+1)\over x}+{4\mu x\over x^2+1}-2wa^2x\right)f_n'(x)+\left[2Ea^2-wa^2(2l+3+4\mu)+{2\mu(2l+3+2wa^2)+2\mu(\mu-1)\over x^2+1}\right]f_n(x)=0.
\end{align}
We now consider the cases where the following two equations are satisfied 
\begin{equation*}
\left\{ \begin{array}{l}
 2\mu(2l+3+2wa^2)+2\mu(\mu-1)=0,\\ \\
  g=\mu(\mu-1).
       \end{array} \right.
\end{equation*}
The solutions of this system, for $g$ and $\mu$, are given explicitly by 
\begin{equation}\label{eq12}
\left\{ \begin{array}{l}
 g=0,\\ \\
 \mu=0,
       \end{array} \right.\quad\mbox{or}\quad\left\{ \begin{array}{l}
 g=2 (1 + l + a^2 w) (3 + 2 l + 2 a^2 w),\\ \\
 \mu=-2 (1+l+a^2 w).
       \end{array} \right.
\end{equation}
In the next, we consider each case of these two sets of solutions. 
\subsection{Case 1}
\noindent The first set of solutions $(g,\mu)=(0,0)$ reduces the differential equation (\ref{eq9}) to
\begin{align}\label{eq13}
xf_n''(x)&+[-2wa^2x^2+2(l+1)]f_n'(x)+(2Ea^2-wa^2(2l+3))~x~f_n(x)=0
\end{align}
which is a special case of the general differential equation
\begin{equation}\label{eq14}
(a_{3,0}x^3+a_{3,1}x^2+a_{3,2}x+a_{3,3})~y^{\prime \prime}+(a_{2,0}x^2+a_{2,1}x+a_{2,2})~y'-(\tau_{1,0} x+\tau_{1,1})~y=0,
\end{equation}
with $a_{3,0}=a_{3,1}=a_{3,3}=a_{2,1}=\tau_{1,1}=0$, $a_{3,2}=1$, $a_{2,0}=-2wa^2,a_{2,2}=2(l+1)$, and $\tau_{1,0}=-2Ea^2+wa^2(2l+3)$. The necessary and sufficient conditions for polynomial solutions of Eq.(\ref{eq14}) are given by the following theorem \cite{C3}.
\vskip0.1true in
\noindent{\bf Theorem 1.} \emph{
The second-order linear differential equation (\ref{eq14}) has a polynomial solution of degree $n$ if 
\begin{equation}\label{eq15}
\tau_{1,0}=n(n-1)a_{3,0}+na_{2,0},\quad n=0,1,2,\dots,
\end{equation}
along with the vanishing of $(n+1)\times(n+1)$-determinant $\Delta_{n+1}$ given by
\begin{center}
$\Delta_{n+1}$~~=~~\begin{tabular}{|lllllll|}
 $\beta_0~~$ & $\alpha_1$ &$\eta_1$&~& ~&~ &~\\
  $\gamma_1$ & $\beta_1$ &  $\alpha_2$&$\eta_2$&~&~&~ \\
~ & $\gamma_2$  & $\beta_2$&$\alpha_3$&$\eta_3$&~&~\\
$~$&~&$\ddots$&$\ddots$&$\ddots$&$\ddots$&~\\ 
~&~&~&$\gamma_{n-2}$&$\beta_{n-2}$&$\alpha_{n-1}$&$\eta_{n-1}$\\
~&~&~&&~$\gamma_{n-1}$&$\beta_{n-1}$&$\alpha_n$\\
~&~&~&~&$~$&$\gamma_{n}$&$\beta_n$\\
\end{tabular}~~=~~0
\end{center}
where
\begin{align}\label{eq16}
\beta_n&=\tau_{1,1}-n((n-1)a_{3,1}+a_{2,1})\notag\\
\alpha_n&=-n((n-1)a_{3,2}+a_{2,2})\notag\\
\gamma_n&=\tau_{1,0}-(n-1)((n-2)a_{3,0}+a_{2,0})\notag\\
\eta_n&=-n(n+1)a_{3,3}
\end{align}
and $\tau_{1,0}$ is fixed for a given $n$ in the determinant $\Delta_{n+1}=0$.
}
\vskip0.1true in

\noindent Thus, the necessary condition for the differential equation (\ref{eq13}) to have polynomial solutions $f_n(x)=\sum_{i=0}^n c_i x^i$ is 
\begin{equation}\label{eq17}
2E_na^2=wa^2(2n'+2l+3),\quad n'=0,1,2,\dots
\end{equation}
while the sufficient condition, Eq(\ref{eq16}), is 
\begin{center}
$\Delta_{n+1}$~~=~~\begin{tabular}{|lllllll|}
 $0~~$ & $\alpha_1$ &0&0& ~&~ &~\\
  $\gamma_1$ & $0$ &  $\alpha_2$&0&~&~&~ \\
~ & $\gamma_2$  &0&$\alpha_3$&0&~&~\\
$~$&~&$\ddots$&$\ddots$&$\ddots$&$\ddots$&~\\ 
~&~&~&$\gamma_{n-2}$&0&$\alpha_{n-1}$&0\\
~&~&~&&~$\gamma_{n-1}$&0&$\alpha_n$\\
~&~&~&~&$~$&$\gamma_{n}$&0\\
\end{tabular}~~=~$\left\{ \begin{array}{ll}
 0 &\mbox{if $n=0,2,4,\dots$} \\ \\
  \prod\limits_{j=0}^{n-1\over 2}(-1)^{2j+1}\alpha_{2j+1}\gamma_{2j+1}=0 &\mbox{if $n=1,3,5,\dots$}
       \end{array} \right.$
\end{center}
where $\beta_n=0$, $\alpha_n=-n(n+2l+1)$ and $\gamma_n=2wa^2(n-n'-1)$.
\vskip0.1true in
\noindent If $l=-1$, the determinant $\Delta_{n+1}$ is identically zero for all $n$, which is equivalent to the exact solutions of the one-dimensional harmonic oscillator problem. 
\vskip0.1true in
\noindent For $l\neq -1$, we have for $n=0,2,4,\dots$, $\Delta_{n+1}\equiv 0$ and we obtain the exact solutions of the Gol'dman and Krivchenkov (or Isotonic) Hamiltonian  $H_0$ where
\begin{equation}\label{eq18}
H_0\psi_{nl}(x)\equiv \left[-{d^2\over dx^2}+{l(l+1)\over x^2}+w^2a^4x^2\right]\psi_{nl}(x)=2E_{nl}^{g=0}a^2\psi_{nl}(x),\quad 0\leq x<\infty.
\end{equation}
These exact solutions are given by \cite{hall1} 
\begin{equation}\label{eq19}
\left\{ \begin{array}{l}
2a^2E_{nl}^{g=0}=wa^2(4n+2l+3), n=0,1,2,\dots\\
\\
 \psi_{nl}(x)= x^{l+1}e^{-wa^2x^2/2}{}_1F_1(-n;l+{3\over 2};wa^2x^2),\quad n=0,1,2,\dots.
       \end{array} \right.
\end{equation}
where the confluent hypergeometric function ${}_1F_1(-n;a;z)$ defined, in terms of the Pochhammer symbol (or Gamma function $\Gamma(a)$) 
$$(a)_k={\Gamma(a+k)\over \Gamma(a)}= \left\{ \begin{array}{ll}
1 &\mbox{if $(k=0, a\in \mathbb C\backslash\{0\})$} \\
  a(a+1)(a+2)\dots(a+k-1) &\mbox{if $(k=\mathbb N, a\in \mathbb C)$}
       \end{array} \right.
$$ 
as
\begin{equation}\label{eq20}
{}_1F_1(-n;a;z)=\sum_{k=0}^n {(-n)_k z^k\over (a)_kk!}.
\end{equation}
The polynomial solutions $f_n(x)={}_1F_1(-n;l+{3\over 2};wa^2x^2)$ are easily obtained by using the asymptotic iteration method (AIM), which is summarized by means of the following theorem.
\vskip0.1true in
\noindent{\bf Theorem 2:} (H. Ciftci et al.\cite{ciftci1}, equations (2.13)-(2.14)) \emph{Given $\lambda_0\equiv\lambda _{0}(x)$ and $s_0\equiv s_{0}(x)$ in $C^{\infty }$, the differential equation 
$$f''(x)=\lambda_0(x)f'(x)+s_0(x)f(x)$$
has the general solution%
\begin{equation}\label{eq21}
f(x)=\exp \left( -\int\limits^{x}\alpha (t)dt\right) \left[ C_{2}+C_{1}\int%
\limits^{x}\exp \left( \int\limits^{t}\left( \lambda _{0}(\tau )+2\alpha
(\tau )\right) d\tau \right) dt\right]
\end{equation}%
if for some $n\in \mathbb N^+=\{1,2,\dots\}$ 
\begin{equation}\label{eq22}
\frac{s_{n}}{\lambda _{n}}=\frac{s_{n-1}}{\lambda _{n-1}}=\alpha (x)
,\quad\mbox{or}\quad\delta _{n}(x)=\lambda _{n}s_{n-1}-\lambda _{n-1}s_{n}=0,
\end{equation}%
where
\begin{align}\label{eq23}
\lambda _{n}&=\lambda _{n-1}^{\prime }+s_{n-1}+\lambda_{0}\lambda _{n},\notag \\
 s_{n}&=s_{n-1}^{\prime }+s_{0}\lambda _{n}.
\end{align}
}
For the differential equation (\ref{eq13}) with
\begin{equation}\label{eq24}
\left\{ \begin{array}{l}
 \lambda_0(x)= -{(-2wa^2x^2+2(l+1))\over x},\\ \\
  s_0(x)=-(2Ea^2-wa^2(2l+3),
       \end{array} \right.
\end{equation}
the first few iterations with $\delta_n=\lambda _{n}s_{n-1}-\lambda _{n-1}s_{n}=0$, using (\ref{eq21}), implies
\begin{equation}\label{eq25}
\left\{ \begin{array}{l}
 f_0(x)=1 \\
  f_1(x)= 2wa^2 x^2-(2l+3)\\
f_2(x)=4w^2a^4x^4-4wa^2(2l+5)x^2+(2l+3)(2l+5)\\
\dots
       \end{array} \right.
\end{equation}
which we may easily generalized using the definition of the confluent hypergeometric function, Eq(\ref{eq20}), as 
\begin{equation}\label{eq26}
f_n(x)={}_1F_1(-n;l+{3\over 2};wa^2x^2)
\end{equation}
 up to a constant.

\subsection{Case 2}
\noindent The second set of solutions  $$(g,\mu)= (2 (1 + l + a^2 w) (3 + 2 l + 2 a^2 w),-2 (1+l+a^2 w))$$ allow us to write the differential equation (\ref{eq9}) as
\begin{align}\label{eq27}
f_n''(x)&+\left({2(l+1)\over x}-{8(l+1+a^2w)x\over x^2+1}-2wa^2x\right)f_n'(x)+\left(2Ea^2+wa^2(6l+5+8wa^2)\right)f_n(x)=0.
\end{align}
A further change of variable $z=x^2+1$ allows us to write the differential equation (\ref{eq27}) as
\begin{align}\label{eq28}
4z(z-1)f''(z)&-\left(4a^2wz^2+2(6l+5+6wa^2)z-16(l+1+wa^2)\right)f'(z)+(2Ea^2+wa^2(6l+5+8wa^2))z~f(z)=0,
\end{align}
Again, Eq.(\ref{eq28}) is a special case of the differential equation (\ref{eq14}) with $a_{3,0}=a_{3,3}=\tau_{1,1}=0$, $a_{3,1}=4,a_{3,2}=-4, a_{2,0}=-4wa^2, a_{2,1}=-2(6l+5+6wa^2), a_{2,2}=16(l+1+wa^2)$ and $\tau_{1,0}=-2Ea^2-wa^2(6l+5+8wa^2)$.
Consequently, the polynomial solutions $f_n(x)$ of (\ref{eq28}) are subject to the following two conditions: the necessary condition (\ref{eq15})  reads
\begin{equation}\label{eq29}
2E_na^2=wa^2(4n'-6l-5-8wa^2),\quad n'=0,1,2,\dots
\end{equation}
and the sufficient condition; namely, the vanishing of the tridiagonal determinant Eq(\ref{eq16}), reads
\begin{center}
$\Delta_{n+1}$~~=~~\begin{tabular}{|lllllll|}
 $\beta_0~~$ & $\alpha_1$ &~&~& ~&~ &~\\
  $\gamma_1$ & $\beta_1$ &  $\alpha_2$&$~$&~&~&~ \\
~ & $\gamma_2$  & $\beta_2$&$\alpha_3$&~&~&~\\
$~$&~&$\ddots$&$\ddots$&$\ddots$&$~$&~\\ 
~&~&~&$\gamma_{n-2}$&$\beta_{n-2}$&$\alpha_{n-1}$&$~$\\
~&~&~&&~$\gamma_{n-1}$&$\beta_{n-1}$&$\alpha_n$\\
~&~&~&~&$~$&$\gamma_{n}$&$\beta_n$\\
\end{tabular}~~=~~0
\end{center}
where
\begin{align}\label{eq30}
\beta_n&=-2n(2n-6l-7-6wa^2)\notag\\
\alpha_n&=4n(n-4l-5-4a^2w)\notag\\
\gamma_n&=4wa^2(n-n'-1)
\end{align}
and $n'=n$ is fixed for the given dimension of the determinant $\Delta_{n+1}$.  From the sufficient condition (\ref{eq30})  we obtain the following conditions on the parameters
\begin{align*}
\Delta_2&=0 \Rightarrow a^2w(l+1+a^2w)=0\\
\Delta_3&=0\Rightarrow a^2w(l+1+a^2w)(1+2l+2a^2w)=0\\
\Delta_4&=0\Rightarrow a^2w(l+1+a^2w)(1+2l+2a^2w)(3(1+6l)+14a^2w)=0\\
\Delta_5&=0\Rightarrow a^2w(l+1+a^2w)(1+2l+2a^2w)(3(6l-1)(6l+1)+4(38l+1)a^2w+44a^4w^2)=0\\
\Delta_6&=0\Rightarrow a^2w(l+1+a^2w)(1+2l+2a^2w)(3(2l-1)(6l-1)(6l+1)+2(208l^2-54l-5)a^2w+200la^4w^2)=0\\
\dots&=\dots
\end{align*}
For a physically meaningful solution we must have $a^2w>0$. This is possible for a very restricted value of the angular momentum number $l$.  Since $\beta_0=0$, we may observe that
\begin{center}
$\Delta_{n+1}$~~=$(l+1+a^2w)(1+2l+2a^2w)\times$~~\begin{tabular}{|lllllll|}
 $\beta_2~~$ & $\alpha_3$ &~&~& ~&~ &~\\
  $\gamma_3$ & $\beta_3$ &  $\alpha_4$&$~$&~&~&~ \\
~ & $\gamma_4$  & $\beta_4$&$\alpha_5$&~&~&~\\
$~$&~&$\ddots$&$\ddots$&$\ddots$&$~$&~\\ 
~&~&~&$\gamma_{n-2}$&$\beta_{n-2}$&$\alpha_{n-1}$&$~$\\
~&~&~&&~$\gamma_{n-1}$&$\beta_{n-1}$&$\alpha_n$\\
~&~&~&~&$~$&$\gamma_{n}$&$\beta_n$\\
\end{tabular}~~=$(l+1+a^2w)(1+2l+2a^2w)\times Q_{n-1}^l(a^2 w)$
\end{center}
where $Q_{n-1}^l(a^2 w)$ are polynomials in the parameter product $a^2 w$. 
\vskip0.1true in
\noindent For physically acceptable solutions, we must have $l=-1$ and the factor $(l+1+a^2w)$ yields $a^2 w=0$, which is not physically acceptable; so we ignore it. The second factor $(1+2l+2a^2w)$ implies a special value of $a^2 w=1/2$, for all $n$, which we will study shortly in full detail. 
Meanwhile, the polynomials $Q_n^l(a^2w)$ 
\begin{equation}\label{eq31}
Q_{n-1}^{l=-1}(a^2w)=\left\{ \begin{array}{ll}
1& \mbox{if $n=2$} \\
14a^2w-15& \mbox{if $n=3$}\\
44a^4w^2-148a^2w+105& \mbox{if $n=4$}\\
200a^4w^2-514a^2w+315& \mbox{if $n=5$}\\
\dots
       \end{array} \right.
\end{equation}
give new values, not reported before, of $a^2 w$ that yield quasi-exact solutions of the Schr\"odinger equation (with one eigenstate)
\begin{align}\label{eq32}
-\psi_{n}''(x)&+\left[(wa^2)^2x^2+4a^2w(1+2a^2w){(x^2-1)\over (x^2+1)^2}\right]\psi_{n}(x)=wa^2(4n+1-8a^2w)\psi_{n}(x)
\end{align}
where
$$\psi_n(x)=(x^2+1)^{-2a^2w}e^{-wa^2x^2/2}f_n(x),$$
and $f_n(x)$ are the solutions of
\begin{align}\label{eq33}
4z(z-1)f''(z)&-\left(4a^2wz^2+2(-1+6wa^2)z-16wa^2\right)f'(z)+4nwa^2z~f(z)=0,\quad z=x^2+1.
\end{align}
For example, $\Delta_4=0$ implies, using (\ref{eq31}), that $a^2w={15\over 14}$, and thus we have for
  \begin{align}\label{eq34}
-\psi_{3}''(x)&+\left[{{225\over 196}} x^2+{660\over 49}{(x^2-1)\over (x^2+1)^2}\right]\psi_{3}(x)={465\over 98}\psi_{3}(x),
\end{align}
the exact solution
$$\psi_3(x)=(x^2+1)^{-{15\over 7}}e^{-{15\over 28}x^2}(45x^6+225x^4+315x^2-49)$$
with a plot of the wave function and potential given in Figure 1. 

\begin{figure}[htbp]
\label{fig1}
\includegraphics[width=3in, height=3in]{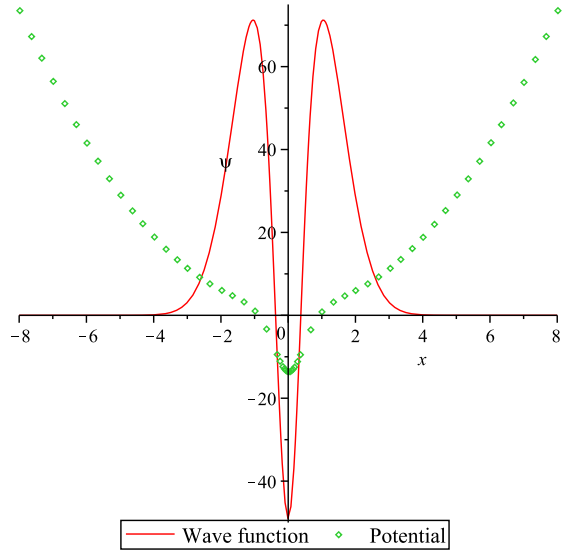}
\caption{Plot of the unnormalized wave function $\psi_3(x)$ and the potential $V_3={{225\over 196}} x^2+{660\over 49}{(x^2-1)\over (x^2+1)^2}$} 
\end{figure}

\noindent Further, $\Delta_5=0$, Eq.(\ref{eq31}) implies
$$a^2w={37\over 22}\pm{\sqrt{214}\over 22}$$
and we have for
\begin{align}\label{eq35}
-\psi_{4}''(x)&+\left[({37\over 22}\pm{\sqrt{214}\over 22})^2x^2+2({37\over 11}\pm{\sqrt{214}\over 11})({48\over 11}\pm{\sqrt{214}\over 11}){(x^2-1)\over (x^2+1)^2}\right]\psi_{4}(x)=({37\over 22}\pm{\sqrt{214}\over 22})({39\over 11}\mp{4\sqrt{214}\over 11})\psi_{4}(x)
\end{align}
the exact solutions
\begin{align*}
\psi_4^{\pm}(x)&=(x^2+1)^{-({37\over 11}\pm{\sqrt{214}\over 11})}e^{-({37\over 44}\pm{\sqrt{214}\over 44})x^2}\\
&(
1575x^8+(9660\pm 420\sqrt{214})x^6+(26250\pm2100\sqrt{214})x^4+(29820\pm2940\sqrt{214})x^2-(1129\pm 188\sqrt{214})).
\end{align*}
Similar results can be obtained for $\Delta_{n+1}=0,$ for $n\geq 5$.
\subsection{Exactly solvable quantum isotonic nonlinear oscillator } 
\noindent As mentioned above, for $l=-1$ and $a^2 w=1/2$, it clear that
$\Delta_{n+1}=0$ for all $n$ and the one-dimensional Schr\"odinger equation
\begin{equation}\label{eq36}
\left[-{d^2\over dx^2}+{x^2\over 4}+{4(x^2-1)\over (x^2+1)^2}\right]\psi_n(x)=(2n-{3\over 2})\psi_n(x),\quad n=0,1,2,\dots
\end{equation}
has the exact solutions
\begin{equation}\label{eq37}
\psi_n(x)=(x^2+1)^{-1}e^{-x^2/4}f_n(x),
\end{equation}
where $f_n(x)$ are the polynomial solutions of the following second-order linear differential equation ($z=x^2+1$)
\begin{align}\label{eq38}
4z(z-1)f_n''(z)&-\left(2z^2+4z-8\right)f_n'(z)+2nz~f_n(z)=0,
\end{align}
By using AIM (Theorem 2, Eq.(\ref{eq21})), we find that the polynomial solutions $f_n(x)$ of Eq.(\ref{eq38}) are given explicitly as
\begin{equation}\label{eq39}
\left\{ \begin{array}{l}
f_0(x)=1\\
f_1(x)= x^2-2\\
f_2(x)=x^3-6x^2+8\\
f_3(x)=x^4-16x^3+52x^2-52\\
f_4(x)=x^5-30x^4+250x^3-580x^2+464\\
\dots
       \end{array} \right.
\end{equation} 
a set of polynomial solutions that can be generated using  
\begin{align}\label{eq40}
f_0(x)=1,\quad f_n(x)&=-3x(2n+1){}_1F_1(-n;{3\over 2};{1\over 2}(x-1))+6((n+1)x-1){}_1F_1(-n+1;{3\over 2};{1\over 2}(x-1)),
\end{align}
up to a constant factor, where, again, ${}_1F_1$ refers to the confluent hypergeometric function defined by (\ref{eq20}). Note that the polynomials $f_n(x)$ in equation (\ref{eq40}) can be expressed in terms of the associated Laguerre polynomials \cite{temme} as
\begin{align}\label{eq41}
f_0(x)=1,f_n(x)&={3(-1)^n\sqrt{\pi}~\Gamma(n)\over 2\Gamma(n+{3\over2})}\bigg[((1+n)(x-1)^2+n)L_n^{{1\over 2}}\left({x-1\over 2}\right)-(x-1)((1+n)x-1)L_n^{{3\over 2}}\left({x-1\over 2}\right)\bigg].
\end{align}
\section{Quasi-polynomial solutions of the generalized quantum isotonic oscillator}
\noindent In this section we study the quasi-polynomial solutions of the differential equation (\ref{eq9}). We note first, using the change of variable $z=x^2$,  Eq.(\ref{eq9}) can be written as
\begin{align}\label{eq42}
&f_n''(z)+\left({2l+3\over 2z}+{2\mu \over z+1}-wa^2\right)f_n'(z)\notag\\
&+\left[{2Ea^2-wa^2(2l+3+4\mu)\over 4z}+{\mu(2l+3+2wa^2)\over 2z(z+1)}-{g\over 2}{(z-1)\over z(z+1)^2}+{\mu(\mu-1)\over (z+1)^2}\right]f_n(z)=0
\end{align}
By means of the M\"obius transformation $z={t/(1-t)}$ that maps the singular points $\{-1,0,\infty\}$ into $\{0,1,\infty\}$, we obtain 
\begin{align}\label{eq43}
f_n''(t)&+\left({2l+3\over 2t(1-t)}+{2(\mu-1) \over 1-t}-{wa^2\over (1-t)^2}\right)f_n'(t)+\left[{\mu(2l+3+2wa^2)\over 2t(1-t)^2}-{g\over 2}{(2t-1)\over t(1-t)^2}+{\mu(\mu-1)\over (1-t)^2}\right]f_n(t)=0,
\end{align}
where we assume 
\begin{equation}\label{eq44}
2Ea^2-(2l+3+4\mu)wa^2=0.
\end{equation}
The differential equation (\ref{eq43}) can be written as
\begin{align}\label{eq45}
(t^3-2t^2+t)f_n''(t)&+\left[-2(\mu-1)t^2+(2\mu-wa^2-l-{7\over 2}) t+{(l+{3\over 2})}\right]f_n'(t)\notag\\
&+\left[(\mu(\mu-1)-{g}) t +{g\over 2}+{\mu(l+{3\over 2}+wa^2)}\right]f_n(t)=0
\end{align}
which we may now compare with equation (\ref{eq14}) in Theorem~1 with $a_{3,0}=1, a_{3,1}=-2,a_{3,2}=1,a_{3,3}=0, a_{2,0}=-2(\mu-1),a_{2,1}=(2\mu-wa^2-l-{7/2}),a_{2,2}={(l+{3/2})},\tau_{1,0}=-(\mu(\mu-1)-{g}),\tau_{1,1}=-{g\over 2}-{\mu(l+{3\over 2}+wa^2)}.$ We, thus, conclude that the quasi-polynomial solutions $f_n(t)$ of Eq.(\ref{eq45}) are subject to the following conditions:
\begin{equation}\label{eq46}
g=(\mu-k)(\mu-k-1),\quad k=0,1,2,\dots
\end{equation}
along with the vanishing of the tridiagonal determinant $\Delta_{n+1}=0$ 
\begin{center}
\begin{tabular}{|llllll|}
 $\beta_0~~$ & $\alpha_1$ &$~~$&~& ~&~ \\
  $\gamma_1$ & $\beta_1$ &  $\alpha_2$&~&~&~ \\
~     & $\gamma_2$  & $\beta_2$&$\alpha_3$&~&~\\
$~$&~&$\ddots$&$\ddots$&$\ddots$&$~$\\ 
~&~&~&$\gamma_{n-1}$&$\beta_{n-1}$&$\alpha_n$\\
~&~&~&~&$\gamma_{n}$&$\beta_n$\\
\end{tabular}~~=~~0
\end{center}
where 
\begin{align}\label{eq47}
 \left\{ \begin{array}{l}
\beta_n=-{1\over 2}(g+(\mu-n)(3+2l+4n+2a^2w)),\\ \\
\alpha_n=-n(n+{l+{1\over 2}}),\\ \\
\gamma_{n}=g-(\mu-n+1)(\mu-n),
       \end{array} \right.
\end{align}
Here, again, $g=(\mu-k)(\mu-k-1)$ is  fixed for given $k=n$, the \emph{fixed} size of the determinant $\Delta_{n+1}$.
\subsection{Particular Case: $n=0$}
\noindent For $k~(fixed)\equiv n=0$, the differential equation (\ref{eq45}) has the exact solution $f_0(t)=1$ if  $g$ and $\mu$ satisfies, simultaneously, the following system of equations
$$g+\mu(3+2l+2a^2w)=0,\quad g=\mu(\mu-1).$$
Solving this system of equations for $g$ and $\mu$, we obtain the following  values of 
\begin{align}\label{eq48}
g=2(1 + l + a^2w)(3 + 2 l + 2 a^2w),\quad \mbox{and}\quad \mu=-2(l+1+wa^2),
\end{align}
and the ground-state energy, in this case, is given by Eq.(\ref{eq44}), namely,
\begin{align}\label{eq49}
Ea^2=-{1\over 2}a^2w(5+6l+8a^2w)
\end{align}
which in complete agreement with the results of Section II.B.
\subsection{Particular Case: $n=1$}
\noindent For $k~(fixed)\equiv n=1$, the determinant $\Delta_2=0$ of (\ref{eq47}) yields
\begin{align}\label{eq50}
\left\{ \begin{array}{l}
 g^2 + g(-1 + 10\mu + 2l(2\mu+1) + 
              2a^2  w(2\mu-1))+ \mu(\mu -1)(15 + 4 l^2 + 8 l (2 + a^2 w) + 
              4 a^2 w (5 + a^2 w))=0, \\ \\
g-(\mu-1)(\mu-2)=0
       \end{array} \right.
\end{align}
where the energy is given by use of Eq.(\ref{eq44}), for the computed values of $\mu$ and $g$, by 
\begin{equation}\label{eq51}
E=(l+{3\over 2}+2\mu)w.
\end{equation}
Further, Eq.(\ref{eq50}) yields the solutions for $l$ as functions of $\mu$ and $a^2w$ 
\begin{equation}\label{eq52}
l={2-(5+4a^2w)\mu-2\mu^2\pm \sqrt{4-4(3+8a^2w)\mu+9\mu^2}\over 4\mu}\ge -1,
\end{equation}
where the energy states are now given by (\ref{eq51}) along with $l$ given by Eq.(\ref{eq52}). We may also note that for 
\begin{equation}\label{eq53}
a^2w={1\over 2}(k+1),\quad k=0,1,2,\dots
\end{equation}
and
\begin{equation}\label{eq54}
a^2E_{\pm}=-{1\over 8\mu}(k+1)\left(-2+(2k+1)\mu-6\mu^2\pm\sqrt{4-4(4k+7)\mu+9\mu^2}\right).
\end{equation}
Further, for $g=(\mu-1)(\mu-2)$, we obtain the un-normalized wave function (see Eq.(\ref{eq8}))
\begin{equation}\label{eq55}
\psi_{1,l}(x)=x^{l+1}(1+x^2)^{\mu-1}e^{-wa^2x^2/2}(1+{1+2l+\mu+2a^2w\over 5+2l+\mu+2a^2w}x^2)
\end{equation}
Thus, we may summarize these results as follows. The exact solutions of the Schr\"odinger equation (\ref{eq7}) are given by Eqs.(\ref{eq54}) and (\ref{eq55}) \emph{only if} $g$ and $\mu$ are the solutions of the system given by Eq.(\ref{eq50}). In Tables \ref{ta1} and \ref{ta2}, we report few  quasi-exact solutions that can be obtained using this  approach.

\begin{table}[!h]
\caption{Conditions on the value of the parameters $g$ and $\mu$ for the quasi-polynomial solutions in the case of $\Delta_2=0$ with different values of $wa^2$ and $l$.}
{\begin{tabular}{@{}lllll@{}} \toprule
$n$ & $l$ & $wa^2$ & Conditions & $E_{n,l}^{wa^2}\equiv E_{n,l}^{wa^2}(\mu,g)$ \\ [0.5ex]
\hline\hline
1 & $-1$ & ${1\over 2}$ &$
\left\{ \begin{array}{l}
 \mu={1\over 3}\left(-3-15A^{-1/3}-{A^{1/3}}\right),\quad\quad A=3(36-\sqrt{961}) \\
g={1\over 9}A^{-2/3}(15+6A^{1/3}+A^{2/3})(15+9A^{1/3}+A^{2/3})
       \end{array} \right.
$ & $E_{1,-1}^{1\over 2}=-w({3\over 2}+{2\over 3}A^{1/3}+10A^{-1/3})$ \\ \\
~ & $~$ & $1$ &$
\left\{ \begin{array}{l}
 \mu={1\over 3}\left(-5-19A^{-1/3}-A^{1/3}\right),\quad\quad A=161-3\sqrt{2118} \\
g={1\over 9}A^{-2/3}(19+8A^{1/3}+A^{2/3})(19+11A^{1/3}+A^{2/3})
       \end{array} \right.
$ & $E_{1,-1}^1=-w({17\over 6}+{2\over 3}A^{1/3}+{38\over 3}A^{-1/3})$ \\ \\
~ & $~$ & ${3\over 2}$ &$
\left\{ \begin{array}{l}
 \mu={1\over 3}\left(-7-25A^{-1/3}-{A^{1/3}}\right),\quad\quad A=199-18\sqrt{74} \\
g={1\over 9}A^{-2/3}(25+10A^{1/3}+A^{2/3})(25+13A^{1/3}+A^{2/3})
       \end{array} \right.
$ & $E_{1,-1}^{3\over 2}=-w({25\over 6}+{2\over 3}A^{1/3}+{50\over 3}A^{-1/3})$ \\ \\
~ & $~$ & $2$ &$
\left\{ \begin{array}{l}
 \mu={1\over 3}\left(-9-33A^{-1/3}-{A^{1/3}}\right),\quad\quad A=3(72-\sqrt{1191}) \\
g={1\over 9}A^{-2/3}(33+12A^{1/3}+A^{2/3})(33+15A^{1/3}+A^{2/3})
       \end{array} \right.
$ & $E_{1,-1}^2=-w({11\over 2}+{2\over 3}A^{1/3}+22A^{-1/3})$ \\ \\
~ & 0 & ${1\over 2}$ & $
\left\{ \begin{array}{l}
 \mu=0 \\
g=2
       \end{array} \right.
$ & $E_{1,0}^{1\over 2}={3\over 2}w$ \\ \\ 
~ & ~ & ~ & $
\left\{ \begin{array}{l}
 \mu=-{1\over 2} (7+\sqrt{17}) \\
g=29+5\sqrt{17}
       \end{array} \right.
$ & $E_{1,0}^{1\over 2}=-{1\over 2}(11+2\sqrt{17})w$ \\ \\
~ & ~ & ~ & $
\left\{ \begin{array}{l}
 \mu=-{1\over 2} (7-\sqrt{17}) \\
g=29-5\sqrt{17}
       \end{array} \right.
$ & $E_{1,0}^{1\over 2}=-{1\over 2}(11-2\sqrt{17})w$ \\ \\ 
~ & ~ & ${1}$ & $
\left\{ \begin{array}{l}
 \mu=-3+B\\
g=\left(-4+B\right)\left(-5+B\right)\\
B={1\over 3} \Re\left({A^{1/3}+33A^{-1/3}}\right),\quad A=-108+3i\sqrt{2697}
       \end{array} \right.
$ & $E_{1,0}^1=-\left({9\over 2}-{2}B\right)w$ \\ \\ 
~ & ~ & ~ & $
\left\{ \begin{array}{l}
 \mu=-3-B,\\
g=\left(5+B\right)\left(4+B)\right)\\
B=\Re\left({11(1+i\sqrt{3})A^{-1/3}\over 2}+{(1-i\sqrt{3})A^{1/3}\over 6}\right),\quad A=-108+3i\sqrt{2697} 
       \end{array} \right.
$ & $E_{1,0}^{1}=-({9\over 2}+2B)w$ \\ \\
~ & ~ & ~ & $
\left\{ \begin{array}{l}
 \mu=-3-B,\\
g=\left(5+B\right)\left(4+B)\right)\\
B=\Re\left({11(1-i\sqrt{3})A^{-1/3}\over 2}+{(1+i\sqrt{3})A^{1/3}\over 6}\right),\quad A=-108+3i\sqrt{2697} 
       \end{array} \right.
$ & $E_{1,0}^{1}=-({9\over 2}+2B)w$ \\ \\
\hline
\hline
\end{tabular} \label{ta1}}
\end{table}

\begin{table}[!ht]
\caption{Conditions on the value of the parameters $g$ and $\mu$ for the quasi-polynomial solutions in the case of $\Delta_2=0$ with different values of $wa^2$ and $l$.}
{\begin{tabular}{@{}lllll@{}} \toprule
$n$ & $l$ & $wa^2$ & Conditions & $E_{n,l}^{wa^2}\equiv E_{n,l}^{wa^2}(\mu,g)$ \\ [0.5ex]
\hline\hline
~ & 0 & ${3\over 2}$ & $
\left\{ \begin{array}{l}
 \mu=-{11\over 3}+B\\
g=\left(-{14\over 3}+B\right)\left(-{17\over 3}+B\right)\\
B={1\over 3} \Re\left({A^{1/3}+43A^{-1/3}}\right),\quad A=-98+9i\sqrt{863}
       \end{array} \right.
$ & $E_{1,0}^{3\over 2}=-{1\over 6}\left({35}-{12}B\right)w$ \\ \\ 
~ & ~ & ~ & $
\left\{ \begin{array}{l}
 \mu=-{11\over 3}-B,\\
g={1\over 9}\left(17+3B\right)\left(14+3B)\right)\\
B={1\over 6}\Re\left(43(1+i\sqrt{3})A^{-1/3}+(1-i\sqrt{3})A^{1/3}\right),\quad A=-98+9i\sqrt{863} 
       \end{array} \right.
$ & $E_{1,0}^{1}=-{1\over 6}(35+12B)w$ \\ \\
~ & ~ & ~ & $
\left\{ \begin{array}{l}
 \mu=-{11\over 3}-B,\\
g={1\over 9}\left(17+3B\right)\left(14+3B)\right)\\
B={1\over 6}\Re\left(43(1-i\sqrt{3})A^{-1/3}+(1+i\sqrt{3})A^{1/3}\right),\quad A=-98+9i\sqrt{863} 
       \end{array} \right.
$ & $E_{1,0}^{1}=-{1\over 6}(35+12B)w$ \\ \\
~ & ~ & $2$ & $
\left\{ \begin{array}{l}
 \mu=-{13\over 3}+B\\
g={1\over 9}\left(-16+3B\right)\left(-{19}+3B\right)\\
B={1\over 3} \Re\left({A^{1/3}+55A^{-1/3}}\right),\quad A=-55+165i\sqrt{6}
       \end{array} \right.
$ & $E_{1,0}^{3\over 2}=-{1\over 6}\left({43}-{12}B\right)w$ \\ \\ 
~ & ~ & ~ & $
\left\{ \begin{array}{l}
 \mu=-{13\over 3}-B,\\
g={1\over 9}\left(16+3B\right)\left(19+3B)\right)\\
B={1\over 6}\Re\left(55(1+i\sqrt{3})A^{-1/3}+(1-i\sqrt{3})A^{1/3}\right),\quad A=-55+165i\sqrt{6} 
       \end{array} \right.
$ & $E_{1,0}^{1}=-{1\over 6}(43+12B)w$ \\ \\
~ & ~ & ~ & $
\left\{ \begin{array}{l}
 \mu=-{13\over 3}-B,\\
g={1\over 9}\left(16+3B\right)\left(19+3B)\right)\\
B={1\over 6}\Re\left(55(1-i\sqrt{3})A^{-1/3}+(1+i\sqrt{3})A^{1/3}\right),\quad A=-55+165i\sqrt{6} 
       \end{array} \right.
$ & $E_{1,0}^{1}=-{1\over 6}(43+12B)w$ \\ \\
\hline
\hline
\end{tabular} \label{ta2}}
\end{table}
\subsection{Particular Case $n=2$}
\noindent For $k~(fixed)\equiv n=2$, the determinant $\Delta_3=0$ along with the necessary condition (\ref{eq47}) yields
\begin{equation}\label{eq56}
\left\{ \begin{array}{l}
 g^3 + 3g^2(7\mu-1+2 l (1+\mu)+2 a^2w (\mu-1))
-g[18+56l+8l^2+18(7+2l)\mu-3(5+2l)(7+2l)\mu^2\\
-12 a^2w (\mu-1) ((7+2 l) \mu-4)-4 a^4 w^2(2+3 (\mu-2) \mu)]
+\mu(\mu-2 ) (\mu-1)(105 + 142 l + 60 l^2 + 8 l^3 + 
   6 a^2 w (5 + 2 l) (7 + 2 l) \\
+ 12 a^4w^2  (7 + 2 l) + 8 a^6 w^3)
=0, \\ \\
g-(\mu-2)(\mu-3)=0
       \end{array} \right.
\end{equation}
where, again, the energy is given, for the computed values of $\mu$ and $g$ using Eqs.(\ref{eq44}) and (\ref{eq56}), by 
$$E=(l+{3\over 2}+2\mu)w.$$ 
In Table \ref{table:quasi}, we report the numerical results for some of the exact solutions of $\mu$ and $g$ using Eq. (\ref{eq56}) and the values of $(l,wa^2)=(-1,{1\over2})$,  $(l,wa^2)=(-1,{1})$, $(l,wa^2)=(-1,{3\over2})$, $(l,wa^2)=(-1,{2})$ $(l,wa^2)=(0,{1\over2})$,  and  $(l,wa^2)=(0,2),$  respectively. We have also computed the corresponding eigenvalues $E_{2,l}^{wa^2}\equiv E_{2,l}^{wa^2}(\mu,g)$. 

\begin{table}[ht]
\caption{Exact eigenvalues for different values of $l$ and $wa^2$ in the case $\Delta_3=0$.}
{\begin{tabular}{@{}lllll@{}} \toprule
$n$ & $l$ & $wa^2$ & Conditions & $E_{n,l}\equiv E_{n,l}^{wa^2}(\mu,g)$ \\ [0.5ex]
\hline\hline
$2$ & $-1$ & ${1\over 2}$ & $\mu_1=-6.301870878994198$ & $E_{2,-1}^{1\over 2}=-6.051870878994198$ \\ 
$~$ & $~$ & $~$ & $g_1=77.22293097048609$ & $~$ \\ 
\hline
~ & ~ & ~ & $\mu_2=-2.4855365082108594$ & $E_{2,-1}^{1\over 2}=-2.2355365082108594$ \\ 
$~$ & $~$ & $~$ & $g_2=24.605574274703333$ & $~$ \\ 
\hline \hline
$~$ & $~$ & $1$ & $\mu_1=-7.398182984326876$ & $E_{2,-1}^1=-7.148182984326876$ \\ 
$~$ & $~$ & $~$ & $g_1=97.7240263912181$ & $~$ \\ 
\hline
~ & ~ & ~ & $\mu_2=-3.3550579014968194$ & $E_{2,-1}^1=-3.1050579014968194$ \\ 
$~$ & $~$ & $~$ & $g_2=34.03170302988033$ & $~$\\ 
\hline
~ & ~ & ~ & $\mu_3=0.9498105417574756$ & $E_{2,-1}^2=1.1998105417574756$ \\ 
$~$ & $~$ & $~$ & $g_3=2.1530873564462514$ & $~$ \\
\hline\hline
$~$ & $~$ & ${3\over 2}$ & $\mu_1=-8.469623341124414$ & $E_{2,-1}^{3\over 2}=-8.219623341124414$ \\ 
$~$ & $~$ & $~$ & $g_1=120.08263624614156$ & $~$ \\ 
\hline
~ & ~ & ~ & $\mu_2=-4.27750521216504$ & $E_{2,-1}^1=-4.02750521216504$ \\ 
$~$ & $~$ & $~$ & $g_2=45.684576900924284$ & $~$\\ 
\hline
~ & ~ & ~ & $\mu_3=0.9282653601757613$ & $E_{2,-1}^1=1.1782653601757613$ \\ 
$~$ & $~$ & $~$ & $g_3=2.2203497780234294$ & $~$ \\
\hline\hline
$~$ & $~$ & $2$ & $\mu_1=-9.525122115065386$ & $E_{2,-1}^2=-9.275122115065383$ \\ 
$~$ & $~$ & $~$ & $g_1=144.35356188223463$ & $~$ \\ 
\hline
~ & ~ & ~ & $\mu_2=-5.226942179911145$ & $E_{2,-1}^2=-4.976942179911145$ \\ 
$~$ & $~$ & $~$ & $g_2=59.45563545168999$ & $~$\\ 
\hline
~ & ~ & ~ & $\mu_3=0.9186508169859244$ & $E_{2,-1}^2=1.1686508169859244$ \\ 
$~$ & $~$ & $~$ & $g_3=2.250665238619284$ & $~$ \\
\hline\hline
$2$ & $0$ & ${1\over 2}$ & $\mu_1=-8.032243023438463$ & $E_{2,-1}^2=-7.282243023438463$ \\ 
$~$ & $~$ & $~$ & $g_1=110.67814310476818$ & $~$ \\ 
\hline
~ & ~ & ~ & $\mu_2=-4.32825470612182$ & $E_{2,-1}=-3.57825470612182$ \\ 
$~$ & $~$ & $~$ & $g_2=46.37506233167478$ & $~$ \\ \hline
$~$ & $~$ & $2$ & $\mu_1=-11.307737259773461$ & $E_{2,-1}^2=-10.557737259773461$ \\ 
$~$ & $~$ & $~$ & $g_1=190.4036082349363$ & $~$ \\ 
\hline
~ & ~ & ~ & $\mu_2=-7.180564905703867$ & $E_{2,-1}^2=-6.430564905703867$ \\ 
$~$ & $~$ & $~$ & $g_2=93.46333689354533$ & $~$\\ \hline
~ & ~ & ~ & $\mu_3=0.9472009101393033$ & $E_{2,-1}^2=1.6972009101393033$ \\ 
$~$ & $~$ & $~$ & $g_3=2.1611850134722084$ & $~$ \\ 
\hline
\hline
\end{tabular}
\label{table:quasi}}
\end{table}
\section{Numerical computation by use of the asymptotic iteration method}
\noindent For the potential parameters $w,a^2$ and $g$, not necessarily obeying the conditions for quasi-polynomial solutions discussed in the previous sections, the asymptotic iteration method can be employed to compute the eigenvalues of Schr\"odinger equation (\ref{eq7})
for arbitrary values $w,a^2$ and $g$.  
The functions $\lambda_0$ and $s_0$, using Eqs.(\ref{eq43}) and (\ref{eq44}), are given by
\begin{equation}\label{eq57}
\left\{ \begin{array}{l}
\lambda_0(t)=-\left({2l+3\over 2t(1-t)}+{2({Ea^2\over 2wa^2}-{2l+3\over 4}-1)\over (1-t)}-{wa^2\over(1-t)^2}\right), \\ \\
  s_0(t)=-\left({({Ea^2\over 2wa^2}-{2l+3\over 4})(2l+3+2wa^2)\over 2t(1-t)^2}-{g\over 2}{(2t-1)\over t(1-t)^2}+{({Ea^2\over 2wa^2}-{2l+3\over 4})({Ea^2\over 2wa^2}-{2l+3\over 4}-1)\over (1-t)^2}\right),
       \end{array} \right.
\end{equation}
where $t\in (0,1)$. The AIM sequence $\lambda_n(x)$ and $s_n(x)$ 
can be calculated iteratively using the iterative sequences (\ref{eq23}). The energy eigenvalues
of the quantum nonlinear isotonic potential (\ref{eq7}) are obtained from the roots of the termination condition (\ref{eq22}). According to the asymptotic iteration method, in particular the study of Brodie \emph{et al} \cite{brodie}, unless the differential equation is exactly solvable, the termination 
condition (\ref{eq22}) produces for each iteration an expression that depends on both $t$ and $E$ (for given values of the parameters $wa^2$, $g$ and $l$).  In such a case, one faces the problem of finding the best possible starting value $t = t_0$ that stabilizes the AIM process \cite{brodie}. Fortunately, since $t\in (0,1)$, the starting value $t_0$ doesn't represent a serious issue in our eigenvalue calculation using (\ref{eq57}) and the termination condition (\ref{eq22}) in contrast to the case of computing the eigenvalues using $\lambda_0(x)$ and $s_0(x)$ as given by, for example, equation (\ref{eq9}), where $x\in(0,\infty)$. In Table \ref{table:num}, we report our numerical results for energies of the four lowest states of the generalized isotonic oscillator of parameters $w$ and $a$ such that $wa^2 = 2$ and for different values of $g$. In this table, we set $l=-1$ for computing the energies $E_0a^2$ and $E_2a^2$, while we put $l=0$ for computing the energies $E_1a^2$ and $E_3a^2$, respectively. For most of these values, the starting value of $t$ is  $t_0=0.5$ and is shifted towards zero as $g$ gets larger in value. For the values of $g$ that admit a quasi-polynomial solution, the number of iteration doesn't exceed \emph{three}. For most of the other values of $g$, the total number of iteration didn't exceed 65. We found that for $wa^2=2$ and the  values of $g$ reported in Table \ref{table:num}, the number of iteration is relatively small compared to the case of $wa^2={1/2}$ and a large value of the parameter $g$. The numerical computations in the present work were done using {\it Maple} version 13 running on an IBM architecture personal computer in a high-precision environment. In order to accelerate our computation we have written our own code for a root-finding algorithm instead of using the default procedure {\tt Solve} of \emph{Maple 13}. These numerical results are  accurate to the number of decimals reported.

\begin{table}[ht]
\caption{Energies of the four lowest states of the generalized isotonic oscillator of parameters $w$
and $a$ given for $l=-1$ as  $wa^2=2$ and for different values of the parameter $g$. The subscript numbers represents the number of iterations used by AIM.}
{\begin{tabular}{@{}llllll@{}} \toprule
$wa^2$&$g~~$ & $E_0a^2~~$ & $E_1a^2~~$ & $E_2a^2~~$ & $E_3a^2$ \\ [0.5ex]
\hline
\hline
$2$&$0.000~01$ & $~~0.999~993~709~536_{(39)}$ & $~2.999~997~742~768_{(25)}$ & $~4.999~998~464~613_{(32)}$ & $~6.999~998~987~906_{(23)}$ \\ 
\hline
$~$&$0.1$ & $~~0.936~865~790~085_{(43)}$ & $~2.977~274~273~728_{(33)}$ & $~4.984~713~354~070_{(45)}$ & $~6.989~892~949~082_{(32)}$ \\ 
\hline
$~$&$1$ & $~~0.349~595~330~721_{(51)}$ & $~2.758~891~177~876_{(36)}$ & $~4.851~946~642~761_{(42)}$ & $~6.900~301~395~128_{(35)}$ \\ 
\hline
$~$&$2$ & $-0.337~237~264~447_{(51)}$ & $~2.487~025~791~777_{(38)}$ & $~4.709~976~255~628_{(42)}$ & $~6.803~992~334~705_{(34)}$ \\ 
\hline
$~$&$5$ & $-2.549~035~191~007_{(53)}$ & $~1.494~183~218~341_{(39)}$ & $~4.268~043~172~724_{(45)}$ & $~6.534~685~249~316_{(35)}$ \\ 
\hline
$~$&$10$ & $-6.529~142~779~202_{(60)}$ & $-0.660~939~314~881_{(40)}$ & $~3.318~493~978~272_{(46)}$ & $~6.100~400~048~017_{(38)}$ \\ 
\hline
$~$&$12$ & $-8.182~546~155~166_{(65)}$ & $-1.659~292~230~771_{(44)}$ & $~2.838~014~627~229_{(48)}$ & $~5.905~881~549~211_{(39)}$ \\ 
\hline
$~$&$50$ & $-41.876~959~736~225_{(37)}$ & $-26.863~072~307~493_{(33)}$ & $-14.310~287~343~156_{(28)}$ & $-4.206~192~073~796_{(31)}$ \\ 
\hline

\hline
\end{tabular}
\label{table:num}}
\end{table}

\section{Conclusion}

\noindent We have provided a detailed solution of the eigenproblem posed by Schr\"odiger's equation with a generalized nonlinear isotonic oscillator potential.  We have presented a method for computing the quasi-polynomial solutions in cases where the potential parameters satisfy certain conditions.  In other more general cases we have used the asymptotic iteration method to find accurate numerical solutions for arbitrary values of the potential parameters $g$, $w$ and $a$. 
\section*{Acknowledgments}
\medskip
\noindent Partial financial support of this work under Grant Nos.
GP249507 and GP3438 from the Natural Sciences and Engineering
Research Council of Canada is gratefully acknowledged by two of us
(respectively NS and RLH).


\end{document}